\documentclass[conference,final,twocolumn,10pt]{IEEEtran} 

\usepackage{cite}

\ifCLASSINFOpdf
  \usepackage[pdftex]{graphicx}
  \usepackage{epstopdf} 
\else
\fi
\usepackage{amsmath}
\usepackage{algorithm}
\usepackage{algorithmic}

  \usepackage[caption=false,font=footnotesize]{subfig}
\usepackage{url}

\usepackage{amsthm}

\theoremstyle{definition}

\theoremstyle{remark}
\newtheorem*{remark}{Takeaways}

\newcommand{\comment}[1]{{}}

\usepackage{amssymb}
\usepackage{xspace}
\usepackage{bm}
\usepackage{bbm}

\newcommand{\mat}[1]{\ensuremath{\mathbf{#1}}\xspace} 
\renewcommand{\vec}[1]{\ensuremath{\mathbf{#1}}\xspace} 

\newcommand{\parens}[1]{\ensuremath{\left(#1\right)}\xspace}
\newcommand{\brackets}[1]{\ensuremath{\left[#1\right]}\xspace}
\newcommand{\braces}[1]{\ensuremath{\left\{#1\right\}}\xspace}
\newcommand{\bars}[1]{\ensuremath{\left\vert#1\right\vert}\xspace}

\newcommand{\complex}{\ensuremath{\mathbb{C}}\xspace}

\newcommand{\setcomplex}{\ensuremath{\complex}}

\newcommand{\setvector}[2]{\ensuremath{#1^{#2 \times 1}}\xspace}

\newcommand{\setvectorcomplex}[1]{\setvector{\setcomplex}{#1}}

\newcommand{\setmatrix}[3]{\ensuremath{#1^{#2 \times #3}}\xspace}

\newcommand{\setmatrixcomplex}[2]{\setmatrix{\setcomplex}{#1}{#2}}

\newcommand{\trans}{\ensuremath{^{\mathrm{T}}}\xspace}

\newcommand{\entry}[2]{\ensuremath{\brackets{#1}_{#2}}\xspace}

\newcommand{\idx}[1]{\ensuremath{^{\parens{#1}}}\xspace}

\newcommand{\thetatx}{\ensuremath{\theta_{\mathrm{tx}}}\xspace}
\newcommand{\phitx}{\ensuremath{\phi_{\mathrm{tx}}}\xspace}

\newcommand{\thetarx}{\ensuremath{\theta_{\mathrm{rx}}}\xspace}
\newcommand{\phirx}{\ensuremath{\phi_{\mathrm{rx}}}\xspace}

\newcommand{\Ntx}{\ensuremath{N_{\mathrm{tx}}}\xspace}
\newcommand{\Nrx}{\ensuremath{N_{\mathrm{rx}}}\xspace}

\def\vw{{\vec{w}}}

\def\mH{{\mat{H}}}

\def\mL{{\mat{L}}}

\usepackage{xspace}
\usepackage[acronym,nogroupskip,nonumberlist,nopostdot]{glossaries}
\makeglossaries

\usepackage{xcolor}

\newacronym{snr}{SNR}{signal-to-noise ratio}
\newacronym{sinr}{SINR}{signal-to-interference-plus-noise ratio}
\newacronym{inr}{INR}{interference-to-noise ratio}
\newacronym{sir}{SIR}{signal-to-interference ratio}
\newacronym{sqr}{SQR}{signal-to-quantization-noise ratio}
\newacronym{sqnr}{SQNR}{signal-to-quantization-plus-noise ratio}
\newacronym{ian}{IAN}{interference as noise}
\newacronym{ber}{BER}{bit error rate}
\newacronym{pn}{PN}{pseudorandom noise}
\newacronym{bfsk}{BFSK}{binary frequency shift keying}
\newacronym{fh}{FH}{frequency-hopped}
\newacronym{fh-bfsk}{FH-BFSK}{frequency-hopped binary frequency shift keying}
\newacronym{crc}{CRC}{cyclic redundancy check}
\newacronym{isi}{ISI}{intersymbol interference}
\newacronym{dsss}{DSSS}{direct-sequence spread spectrum}
\newacronym{ofdm}{OFDM}{orthogonal frequency-division multiplexing}
\newacronym{ofdma}{OFDMA}{orthogonal frequency-division multiple access}
\newacronym{sdr}{SDR}{software-defined radio}
\newacronym{tx}{TX}{transmitter}
\newacronym{rx}{RX}{receiver}
\newacronym{fdd}{FDD}{frequency-division duplexing}
\newacronym{tdd}{TDD}{time-division duplexing}
\newacronym{fdma}{FDMA}{frequency-division multiple access}
\newacronym{tdma}{TDMA}{time-division multiple access}
\newacronym{sdma}{SDMA}{space-division multiple access}
\newacronym[plural=MPCs]{mpc}{MPC}{multipath component}
\newacronym{mui}{MUI}{multi-user interference}

\newacronym{qam}{QAM}{quadrature amplitude modulation}
\newacronym{mqam}{MQAM}{M-ary quadrature amplitude modulation}

\newacronym{ls}{LS}{least-squares}
\newacronym{lms}{LMS}{least mean squares}
\newacronym{rls}{RLS}{recursive least-squares}
\newacronym{rzf}{RZF}{regularized zero-forcing}
\newacronym{mmse}{MMSE}{minimum mean square error}
\newacronym{lmmse}{LMMSE}{linear minimum mean square error}
\newacronym{mse}{MSE}{mean square error}
\newacronym{fft}{FFT}{fast Fourier transform}
\newacronym{dft}{DFT}{discrete Fourier transform}
\newacronym{dtft}{DTFT}{discrete-time Fourier transform}
\newacronym{ctft}{CTFT}{continuous-time Fourier transform}
\newacronym{ml}{ML}{machine learning}
\newacronym[plural=NNs]{nn}{NN}{neural network}
\newacronym[plural=RNNs]{rnn}{RNN}{recurrent neural network}
\newacronym[plural=ADCs]{adc}{ADC}{analog-to-digital converter}
\newacronym[plural=DACs]{dac}{DAC}{digital-to-analog converter}
\newacronym[plural=FPGAs]{fpga}{FPGA}{field-programmable gate array}
\newacronym{evm}{EVM}{error vector magnitude}
\newacronym{enob}{ENOB}{effective number of bits}
\newacronym{zf}{ZF}{zero-forcing}
\newacronym{rv}{r.v.}{random variable}
\newacronym{omp}{OMP}{orthogonal matching pursuit}
\newacronym{svd}{SVD}{singular value decomposition}

\newacronym{sdp}{SDP}{semidefinite programming}
\newacronym{psd}{PSD}{positive semidefinite}
\newacronym{nsd}{NSD}{negative semidefinite}

\newacronym{agc}{AGC}{automatic gain control}
\newacronym{rf}{RF}{radio frequency}
\newacronym{if}{IF}{intermediate frequency}
\newacronym{los}{LOS}{line-of-sight}
\newacronym{nlos}{NLOS}{non-line-of-sight}
\newacronym{ple}{PLE}{path loss exponent}
\newacronym[plural=dB,firstplural=decibels (dB)]{db}{dB}{decibel}
\newacronym[plural=dBm,firstplural=decibel milliwatts (dBm)]{dbm}{dBm}{decibel milliwatts}
\newacronym{pa}{PA}{power amplifier}
\newacronym{lna}{LNA}{low noise amplifier}
\newacronym{cw}{CW}{continuous wave}
\newacronym{papr}{PAPR}{peak-to-average power ratio}
\newacronym{usrp}{USRP}{Universal Software Radio Peripheral}
\newacronym{irr}{IRR}{image rejection ratio}
\newacronym{lo}{LO}{local oscillator}
\newacronym{vm}{VM}{vector modulator}
\newacronym{mmwave}{mmWave}{millimeter wave}
\newacronym{eirp}{EIRP}{effective isotropic radiated power}

\newacronym{csma}{CSMA}{carrier-sense multiple access}
\newacronym{csmaca}{CSMA/CA}{carrier-sense multiple access with collision avoidance}
\newacronym{csmacd}{CSMA/CD}{carrier-sense multiple access with collision detection}
\newacronym{mac}{MAC}{medium access control}
\newacronym{phy}{PHY}{physical layer}
\newacronym{4g}{4G}{fourth generation}
\newacronym{lte}{LTE}{Long-Term Evolution}
\newacronym{4glte}{4G LTE}{\gls{4g} \gls{lte}}
\newacronym{5g}{5G}{fifth generation}
\newacronym{nr}{NR}{New Radio}
\newacronym{5gnr}{5G NR}{5G New Radio}
\newacronym{ieee}{IEEE}{Institute of Electrical and Electronics Engineers}
\newacronym{wifi}{Wi-Fi}{IEEE 802.11}
\newacronym{lan}{LAN}{local area network}
\newacronym{wlan}{WLAN}{wireless local area network}
\newacronym[plural=BSs]{bs}{BS}{base station}
\newacronym[plural=SBSs]{sbs}{SBS}{small-cell base station}
\newacronym[plural=FD-SBSs]{fdsbs}{FD-SBS}{\gls{fd}-enabled \gls{sbs}}
\newacronym[plural=MBSs]{mbs}{MBS}{macrocell base station}
\newacronym[plural=UEs]{ue}{UE}{user equipment}
\newacronym{ul}{UL}{uplink}
\newacronym{dl}{DL}{downlink}
\newacronym{qos}{QoS}{Quality of Service}
\newacronym{fcc}{FCC}{Federal Communications Commission}
\newacronym{iab}{IAB}{integrated access and backhaul}
\newacronym{fab}{FAB}{fixed access and backhaul}
\newacronym{hetnet}{HetNet}{heterogeneous network}
\newacronym{gnb}{gNB}{gNB}

\newacronym{siso}{SISO}{single-input single-output}
\newacronym{mimo}{MIMO}{multiple-input multiple-output}
\newacronym{sumimo}{SU-MIMO}{single-user \gls{mimo}}
\newacronym{mumimo}{MU-MIMO}{multi-user \gls{mimo}}
\newacronym{bf}{BF}{beamforming}
\newacronym{ca}{CA}{constant amplitude}
\newacronym{ula}{ULA}{uniform linear array}
\newacronym{upa}{UPA}{uniform planar array}
\newacronym[\glslongpluralkey={angles of arrival}]{aoa}{AoA}{angle of arrival}
\newacronym[\glslongpluralkey={angles of departure}]{aod}{AoD}{angle of departure}
\newacronym{dof}{DoF}{degrees of freedom}
\newacronym{csi}{CSI}{channel state information}
\newacronym{csit}{CSIT}{\gls{csi} at the transmitter}
\newacronym{csir}{CSIR}{\gls{csi} at the receiver}
\newacronym{cs}{CS}{compressed sensing}

\newacronym{fd}{FD}{in-band full-duplex}
\newacronym{hd}{HD}{half-duplex}
\newacronym{si}{SI}{self-interference}
\newacronym{sic}{SIC}{self-interference cancellation}
\newacronym{soi}{SoI}{signal of interest}
\newacronym{asic}{A-SIC}{analog \acrlong{sic}}
\newacronym{dsic}{D-SIC}{digital \gls{sic}}
\newacronym{star}{STAR}{simultaneous transmit and receive}
\newacronym{warp}{WARP}{Wireless Open-Access Research Platform}
\newacronym{bfc}{BFC}{beamforming cancellation}
\newacronym{ipi}{IPI}{inter-panel-interference}
\newacronym{ipic}{IPIC}{inter-panel-interference cancellation}

\newacronym{qcqp}{QCQP}{quadratically-constrained quadratic programming}
\newacronym{cdf}{CDF}{cumulative density function}

\newacronym{elf}{ELF}{extremely low frequency}
\newacronym{slf}{SLF}{super low frequency}
\newacronym{ulf}{ULF}{ultra low frequency}
\newacronym{vlf}{VLF}{very low frequency}
\newacronym{lf}{LF}{low frequency}
\newacronym{mf}{MF}{medium frequency}
\newacronym{hf}{HF}{high frequency}
\newacronym{vhf}{VHF}{very high frequency}
\newacronym{uhf}{UHF}{ultra high frequency}
\newacronym{shf}{SHF}{super high frequency}
\newacronym{ehf}{EHF}{extremely high frequency}
\newacronym{thf}{THF}{tremendously high frequency}

\newacronym{wncg}{WNCG}{Wireless Networking and Communications Group}
\newacronym{linc}{LINC}{Laboratory of Informatics, Networks, and Communications}
\newacronym{ut}{UT Austin}{The University of Texas at Austin}
\newacronym{uiuc}{UIUC}{University of Illinois at Urbana-Champaign}
\newacronym{usc}{USC}{University of Southern California}
\newacronym{mit}{MIT}{Massachusetts Institute of Technology}
\newacronym{berkeley}{UC Berkeley}{University of California, Berkeley}
\newacronym{osu}{OSU}{Ohio State University}

\newcommand{\upa}{\gls{upa}\xspace}
\newcommand{\upas}{\glspl{upa}\xspace}
\newcommand{\sdr}{\gls{sdr}\xspace}
\newcommand{\sdrs}{\glspl{sdr}\xspace}

\newacronym{iid}{i.i.d.}{independently and identically distributed}

\newcommand{\mmwave}{\gls{mmwave}\xspace}
\newcommand{\mimo}{\gls{mimo}\xspace}

\newcommand{\iab}{\gls{iab}\xspace}

\newcommand{\figref}[1]{\figurename~\ref{#1}}

\usepackage{multirow}

\begin{document}

%
\title{28 GHz Phased Array-Based Self-Interference Measurements for Millimeter Wave Full-Duplex}

%
%
%

\author{
	\IEEEauthorblockN{Aditya Chopra\IEEEauthorrefmark{1}, Ian P.~Roberts\IEEEauthorrefmark{2}, Thomas Novlan\IEEEauthorrefmark{1}, and Jeffrey G.~Andrews\IEEEauthorrefmark{2}}
	\IEEEauthorblockA{\IEEEauthorrefmark{1}AT\&T Labs, Austin, TX, USA.}
	\IEEEauthorblockA{\IEEEauthorrefmark{2}University of Texas at Austin, Austin, TX, USA.}
}

\maketitle




\begin{abstract}
We present measurements of the 28 GHz self-interference channel for full-duplex sectorized multi-panel \mmwave systems, such as \acrlong{iab}.
We measure the isolation between the input of a transmitting phased array panel and the output of a co-located receiving phased array panel, each of which is electronically steered across a number of directions in azimuth and elevation.
In total, nearly 6.5 million measurements were taken in an anechoic chamber to densely inspect the directional nature of the coupling between 256-element phased arrays.
We observe that highly directional \mmwave beams do not necessarily offer widespread high isolation between transmitting and receiving arrays.
Rather, our measurements indicate that steering the transmitter or receiver away from the other \textit{tends} to offer higher isolation but even slight steering changes can lead to drastic variations in isolation.
These measurements can be useful references when developing \mmwave full-duplex solutions and can motivate a variety of future topics including beam/user selection and beamforming codebook design.
\end{abstract}





\glsresetall


\section{Introduction} \label{sec:introduction}
\Gls{mmwave} communication systems equipped with full-duplex capability could offer networks with improvements at the physical layer and in medium access \cite{roberts_wcm}.
In particular, full-duplex \iab could simultaneously deliver access and maintain backhaul in-band, making better use of \mmwave spectrum while reducing latency between the network core and its edge \cite{3GPP_IAB}.
For example, a pole-mounted \iab node with full-duplex capability could serve a user on the ground while simultaneously maintaining wireless backhaul with a fiber-connected node, all using the same \mmwave spectrum.

To enable such a capability, researchers have proposed a number of beamforming-based solutions that leverage dense \mmwave antenna arrays to strategically steer transmission and reception to mitigate self-interference \cite{liu_beamforming_2016,xia_2017,satyanarayana_hybrid_2019,roberts_bflrdr,lopez-valcarce_beamformer_2019,cai_robust_2019}.
The majority of existing proposed solutions, however, have been evaluated using highly idealized \mmwave self-interference channel models that have not yet been verified with measurements.
As a result, the practical efficacy of these solutions remains unknown.

An early attempt to characterize the self-interference channel at \mmwave was in \cite{rajagopal_2014}, where a beam-sweeping approach was taken to measure the received self-interference power for a combination of transmit and receive beams.
A relatively low number of beams were swept using 28 GHz 8 $\times$ 8 \upas in sample indoor and outdoor environments.
While not particularly extensive in characterizing the spatial characteristics nor the distribution of isolation levels, this work provided a valuable first look at the expected isolation levels seen by a multi-panel \mmwave full-duplex system.
In \cite{kohda_2015}, a proof-of-concept 60 GHz, short-range, full-duplex link was established using phased array transceivers, which found that there was a clear performance improvement when the angle difference between the two transceivers was varied.
The work of \cite{lee_2015} presents indoor and outdoor self-interference channel measurements at approximately 28 GHz using a pair of horn antennas for transmission and reception as well as dipole antennas.
Little was shown on the impacts and variability due to transmit and receive steering direction, though it was noted that some steering combinations were starkly more favorable than others.
In \cite{yang_2016} and \cite{he_2017}, the authors conducted indoor measurements of the 60 GHz self-interference channel with a rotating channel sounder using fixed horn antennas, which showed large variations in self-interference power across azimuth and elevation due to reflective objects such as furniture.

The data and insights on the \mmwave self-interference channel in \cite{rajagopal_2014,kohda_2015,lee_2015,yang_2016,he_2017,haneda_2018} are certainly useful, but do not provide much in terms of proposing or verifying \mimo channel models that \mmwave full-duplex research relies so heavily on.
To evaluate beamforming-based \mmwave full-duplex solutions thus far, researchers have primarily used relatively idealized models, particularly the spherical-wave \mimo model of \cite{spherical_2005}.
The extremely idealized geometric model in \cite{spherical_2005} is sensitive to small modeling errors and does not capture the isolation and effects of practical systems such as enclosures, mounting infrastructure, and non-isotropic antenna elements.
To account for environmental reflections, which were observed in \cite{rajagopal_2014}, it is common for researchers to combine a ray-based model with the spherical-wave model in a Rician fashion \cite{li_2014,satyanarayana_hybrid_2019}.
These channel models have been widely used by \mmwave full-duplex literature but have yet to be verified with measurement.

In this work, we conduct the first known \textit{dense, phased array-based} measurements of the 28 GHz self-interference channel.
We use 256-element \upas to take nearly 6.5 million measurements of the self-interference channel using finely steered, highly directional beams.
In an anechoic chamber, we measure the isolation between each pair of transmit and receive beams to better understand the spatial profile of the direct coupling between a transmitting and receiving arrays of a full-duplex transceiver.
Our measurements show that:
\begin{itemize}
    \item generally, more isolation is seen when the transmitter or receiver steer away from the other
    \item small changes in transmit and receive steering directions can lead to significant changes in  isolation.
\end{itemize}
These measurements act as a first step toward practical characterization of \mmwave self-interference and shed light on the levels of isolation one can realistically expect between transmit and receive beams of a \mmwave full-duplex transceiver.
Our future work will involve modeling \mmwave self-interference.



\section{Measurement Setup \& Methodology} \label{sec:setup}


\begin{figure}
    \centering
    \includegraphics[width=\linewidth,height=\textheight,keepaspectratio]{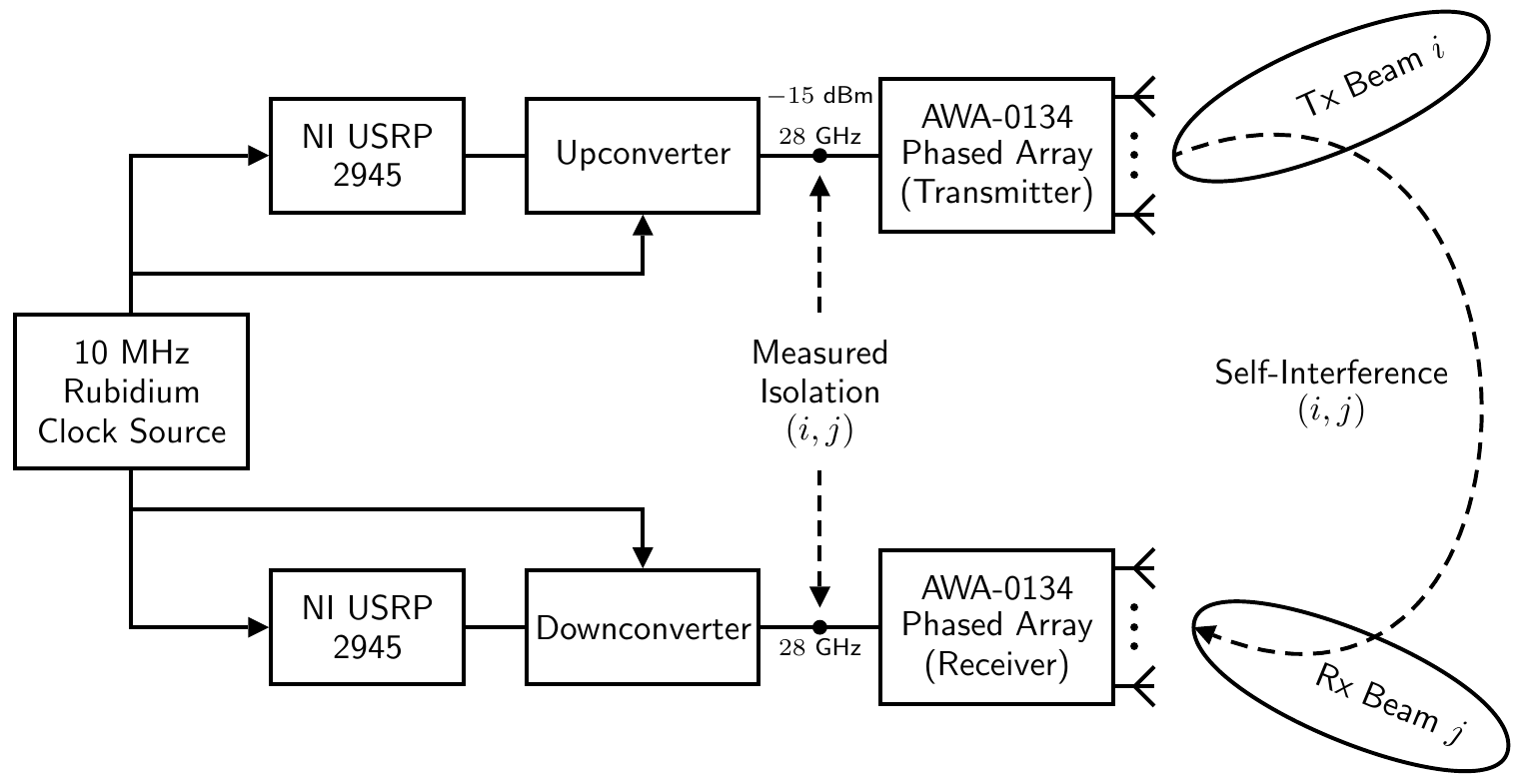}
    \caption{A simplified block diagram of our measurement setup.}
    \label{fig:setup}
\end{figure}

\begin{figure}
    \centering
    \includegraphics[width=\linewidth,keepaspectratio]{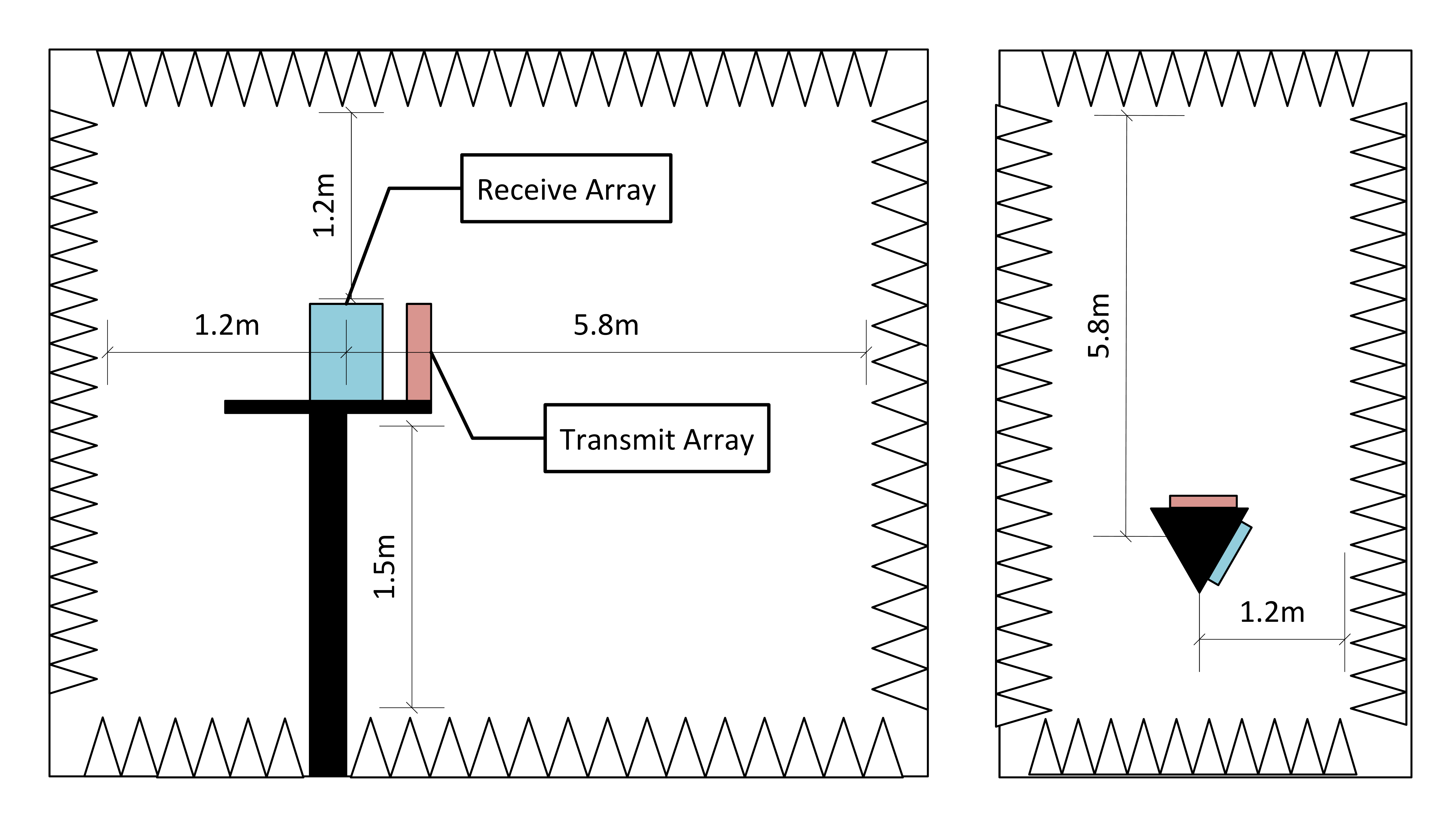}
    \caption{A side-view (left) and top-view (right) of the phased array measurement platform in the anechoic chamber.}
    \label{fig:chamber}
\end{figure}

Our measurement setup, illustrated as a functional block diagram in \figref{fig:setup}, consists of two identical Anokiwave AWA-0134 28 GHz 256-Element Active Antenna Innovator's Kit phased array modules \cite{anokiwave}: one for transmission and one for reception. 
Each of these modules consists of a $16 \times 16$ half-wavelength \upa designed for the 26.5 GHz to 29.5 GHz frequency band. 
As shown in \figref{fig:chamber}, the array modules are mounted on two faces of a triangular metal platform where each side is $35$ cm in length and each face is separated by an angle of $60^\circ$; this is a realistic sectorized deployment configuration. 
The size of each array module is $26$ cm by $14$ cm by $6$ cm, and the antenna elements are oriented in the center of the array module face.
The centers of the transmit \upa and receive \upa are separated by approximately $30$ cm.

The transmit and receive arrays are each connected to a National Instruments' Universal Software Radio Platform (USRP) 2954 \sdr. 
Up to 120 dB of \mmwave isolation, as defined in \figref{fig:setup}, can be measured by this system to within 0.25 dB of accuracy. There are many design decisions behind achieving such a high level of isolation measurement such as using physically separate transmit and receive baseband \sdrs, driving both \sdrs with a single Rubidium oscillator, and employing novel multi-stage averaging algorithms to reduce noise. 


\comment{
We employed a novel measurement algorithm to reliably measure the received power over a wide range.
We transmit a pseudorandom Zadoff-Chu sequence over 50 MHz of bandwidth at the transmitter. 
Once the receive signal is digitized, we apply a \gls{fft} correlation to this captured signal and estimate the total energy. 
A noise threshold is set in place and only the samples of the correlated output with  magnitude higher than this threshold contribute to the final energy calculation. 
Since the arrays were placed in a fixed location in an anechoic chamber, it was assumed and observed that the isolation levels do not significantly vary over time, allowing us to improve our isolation measurements using averaging. 
To ensure phase noise did not accumulate prohibitively during these averages, conducted two-staged averaging and  discarded phase information when taking executing second-stage averaging.
We chose our two-staged averaging window sizes to counteract this interplay between Gaussian and phase noise.
Using calibrated equipment, we found our approach offered much more reliable power measurements over a wide range of power levels.
}


\begin{figure}
    \centering
    \includegraphics[width=\linewidth,height=\textheight,keepaspectratio]{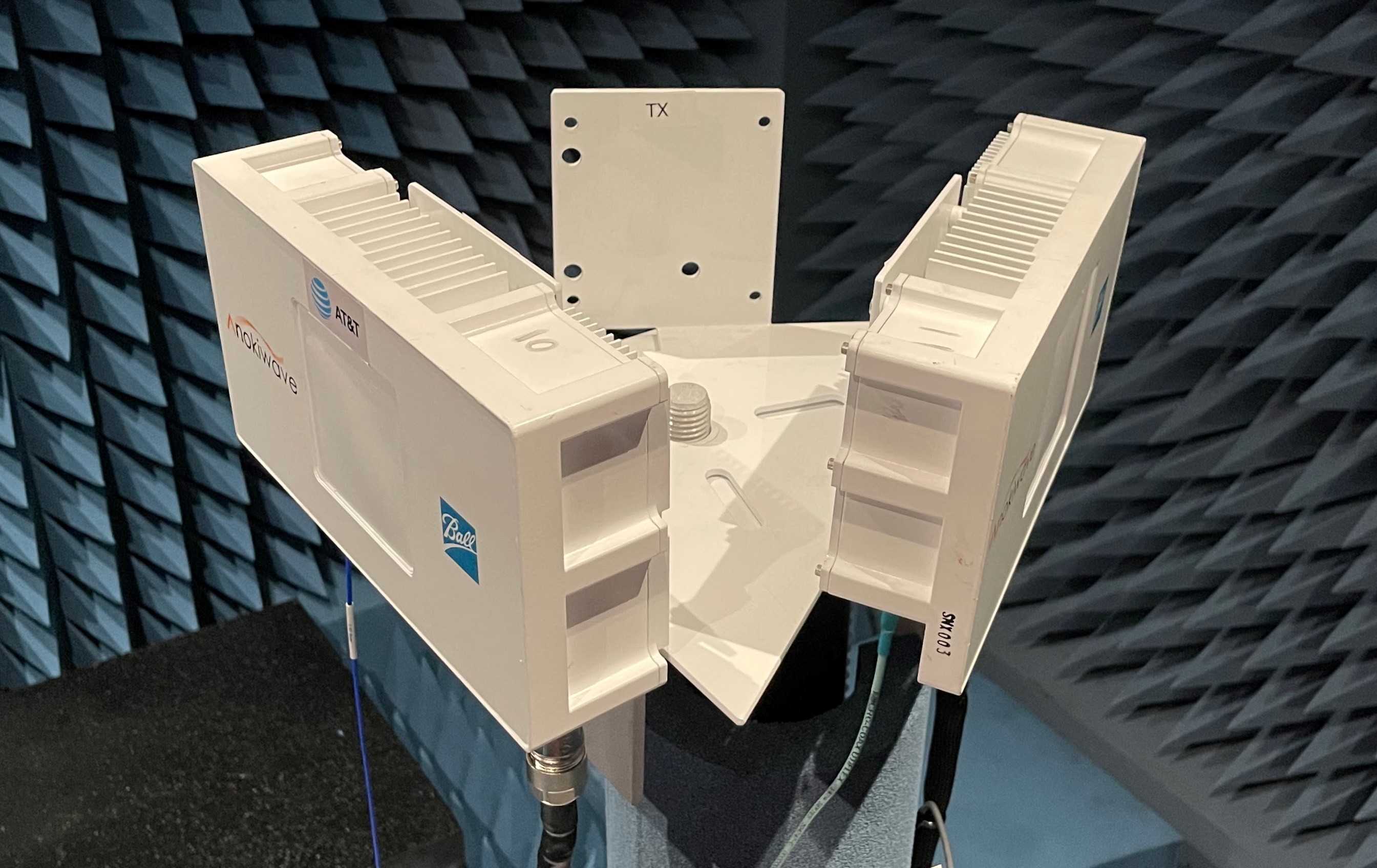}
    \caption{Phased array measurement platform inside an anechoic chamber; receive array on left and transmit array on right.}
    \label{fig:picture}
\end{figure}

The self-interference measurement platform was placed inside an anechoic chamber free from any significant reflectors; future work will investigate the impact of reflectors. The dimensions of the anechoic chamber and the placement of the array inside the chamber are shown using side- and top-views in \figref{fig:chamber}.
A photo of the phased array measurement platform inside the chamber is shown in \figref{fig:picture}. 
Once injected into the phased array, the transmitted signals propagate over the air to the receive array. 
Both the transmit and receive arrays are configured to steer in a desired direction via digitally controlled beamforming weights. 

\begin{figure}
    \centering
    \includegraphics[width=0.89\linewidth,height=\textheight,keepaspectratio]{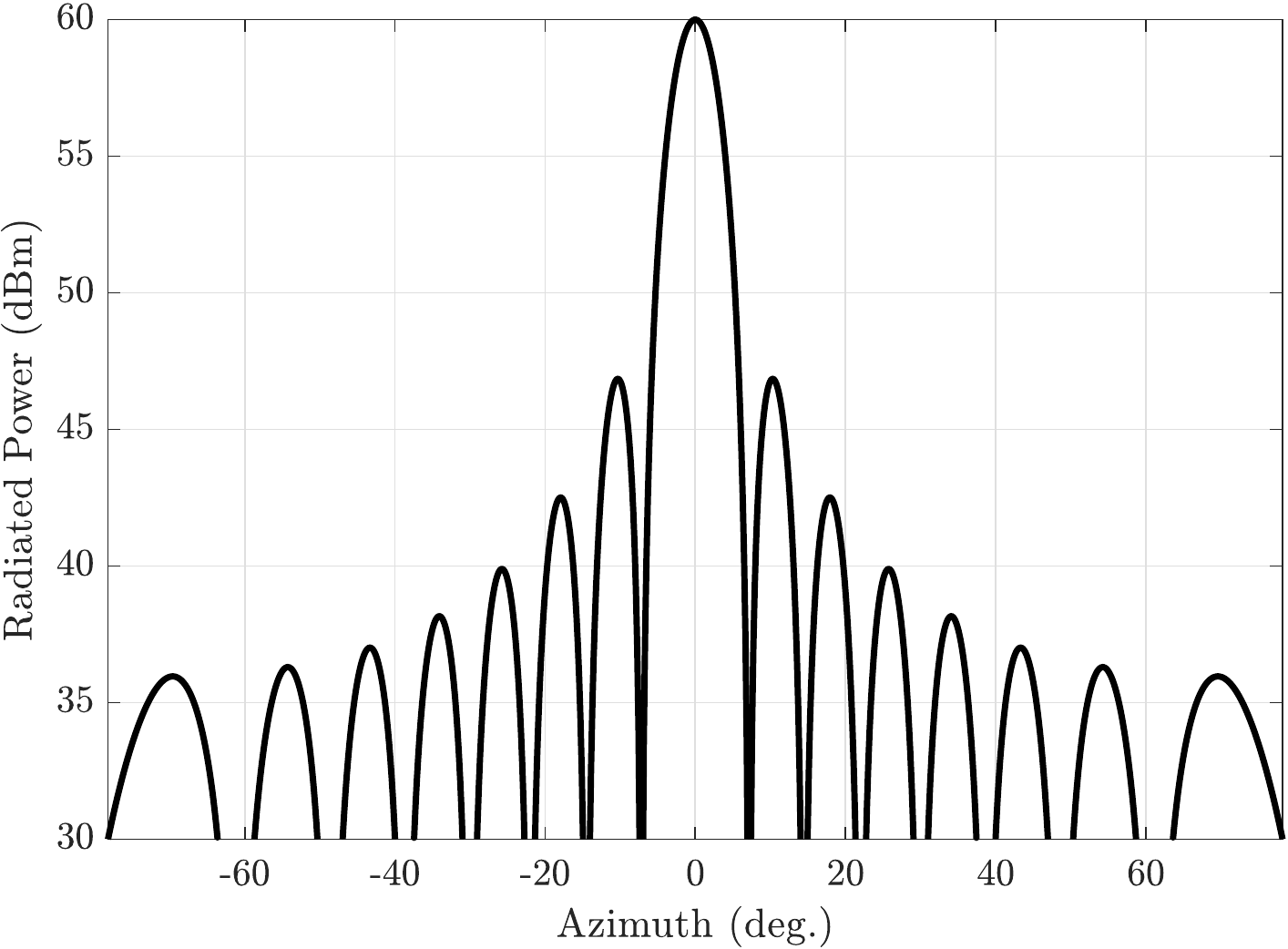}
    \caption{The approximate azimuth radiation pattern of our transmitting $16 \times 16$ \upa steered broadside. The elevation pattern is approximately identical. The \gls{eirp} is 60 dBm and the 3 dB beamwidth is roughly $7^\circ$.}
    \label{fig:beam-pattern}
\end{figure}

For context, the approximate pattern of a beam steered broadside by our transmit array is shown in \figref{fig:beam-pattern}, which delivers an \gls{eirp} of approximately $60$ dBm and has a $3$ dB beamwidth of approximately $7^\circ$ in both azimuth and elevation.
Given that we are interested in measuring isolation between the transmit array and receive array for a variety of transmit and receive beams, it is especially important to take note of what we reference our received signal power to.
Our isolation measurements are captured at the output of the receive array and referenced to the power delivered to the input of the transmit array, which is approximately $-15$ dBm, meaning a measured isolation value of $L$ dB maps to received power of $(-15- L)$ dBm.
As a result, our isolation measurements inherently include gain/losses incurred in between, most notably: (i) spatial combining between the transmit beam, self-interference channel, and receive beam; (ii) amplification in the transmit array module; (iii) amplification in the receive array module; and (iv) various artifacts within the transmit and receive array modules.
While the spatial combining is the primary interest, we have decided not to estimate and abstract out the gains/losses of the other contributors to preserve the accuracy of our measurements.
As such, it is also important to note that the noise power out of our receive array is approximately $-68$ dBm over 100 MHz ($53$ dB below the power delivered into our transmit array).

\section{Summary of Measurements} \label{sec:summary}


To describe the arrays' steering directions in 3-D, we use an azimuth-elevation convention.
A vector in the direction $(\theta,\phi)$ from the center of an array is described by azimuth $\theta$ being the angle between the positive $y$ axis and the vector's orthogonal projection onto the $x$-$y$ plane and elevation $\phi$ being the angle from the $x$-$y$ plane to the vector itself.
Each phased array can be (independently) steered toward a desired direction $(\theta,\phi)$ via beamforming weights $\vw(\theta,\phi) \in \setvectorcomplex{256}$.

A set of $\Ntx$ transmit directions and a set of $\Nrx$ receive directions described as
\begin{gather}
\braces{\parens{\thetatx\idx{n},\phitx\idx{n}}}_{n=1}^{\Ntx}, \quad \braces{\parens{\thetarx\idx{n},\phirx\idx{n}}}_{n=1}^{\Nrx}
\end{gather} 
are specified prior to executing measurement.
The isolation between each combination of transmit-receive beam pairs is measured to form an $\Nrx \times \Ntx$ matrix $\mL$, where
the isolation between the $m$-th transmit beam and the $n$-th receive beam is
\begin{align}
\entry{\mL}{n,m} = \frac{1}{\bars{\vw\parens{\thetarx\idx{n},\phirx\idx{n}}\trans \times \mH \times \vw\parens{\thetatx\idx{m},\phitx\idx{m}}}^2}
\end{align}
where $\mH \in \setmatrixcomplex{256}{256}$ is the over-the-air self-interference channel matrix between our arrays.

\begin{figure}[!t]
    \centering
    \includegraphics[width=\linewidth,height=\textheight,keepaspectratio]{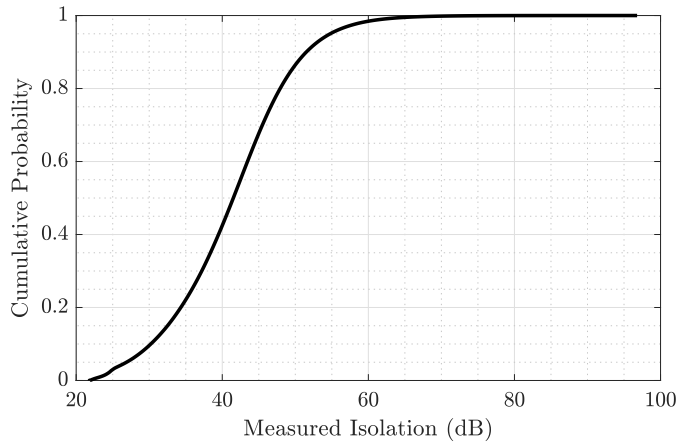}
    \caption{The \gls{cdf} of the isolation measured between nearly 6.5 million pairs of transmit and receive beams.}
    \label{fig:dist-full-cdf}
\end{figure}

The results in this work are based on measurements whose transmit beams and receive beams are each distributed uniformly in azimuth from $-60^\circ$ to $60^\circ$ with $1^\circ$ spacing and in elevation from $-10^\circ$ to $10^\circ$ with $1^\circ$ spacing.
This amounts to a total of $\Ntx = \Nrx = 121 \times 21 = 2541$ beams for transmission and for reception.
Measuring all transmit-receive pairs yields a $2541 \times 2541$ matrix $\mL$ of $6,456,681$ isolation values.
The maximum and minimum isolation seen across transmit-receive beam pairs is $96.79$ dB and $21.81$ dB, respectively.
The mean and median isolation seen across transmit-receive beam pairs is $52.42$ dB and $41.52$ dB, respectively, indicating the distribution of measured isolation values is likely right-skewed, pulling our mean higher than the median.
The \gls{cdf} of measurements, shown in \figref{fig:dist-full-cdf}, confirms this, which is even more apparent when isolation (the $x$ axis) is in linear units.
{We see that high-isolation measurements (e.g., $55$ dB or more) are far less frequent than low-isolation ones.}

\begin{figure*}[!t]
    \centering
    \subfloat[Observed by each transmit beam.]{\includegraphics[width=0.44\linewidth,height=\textheight,keepaspectratio]{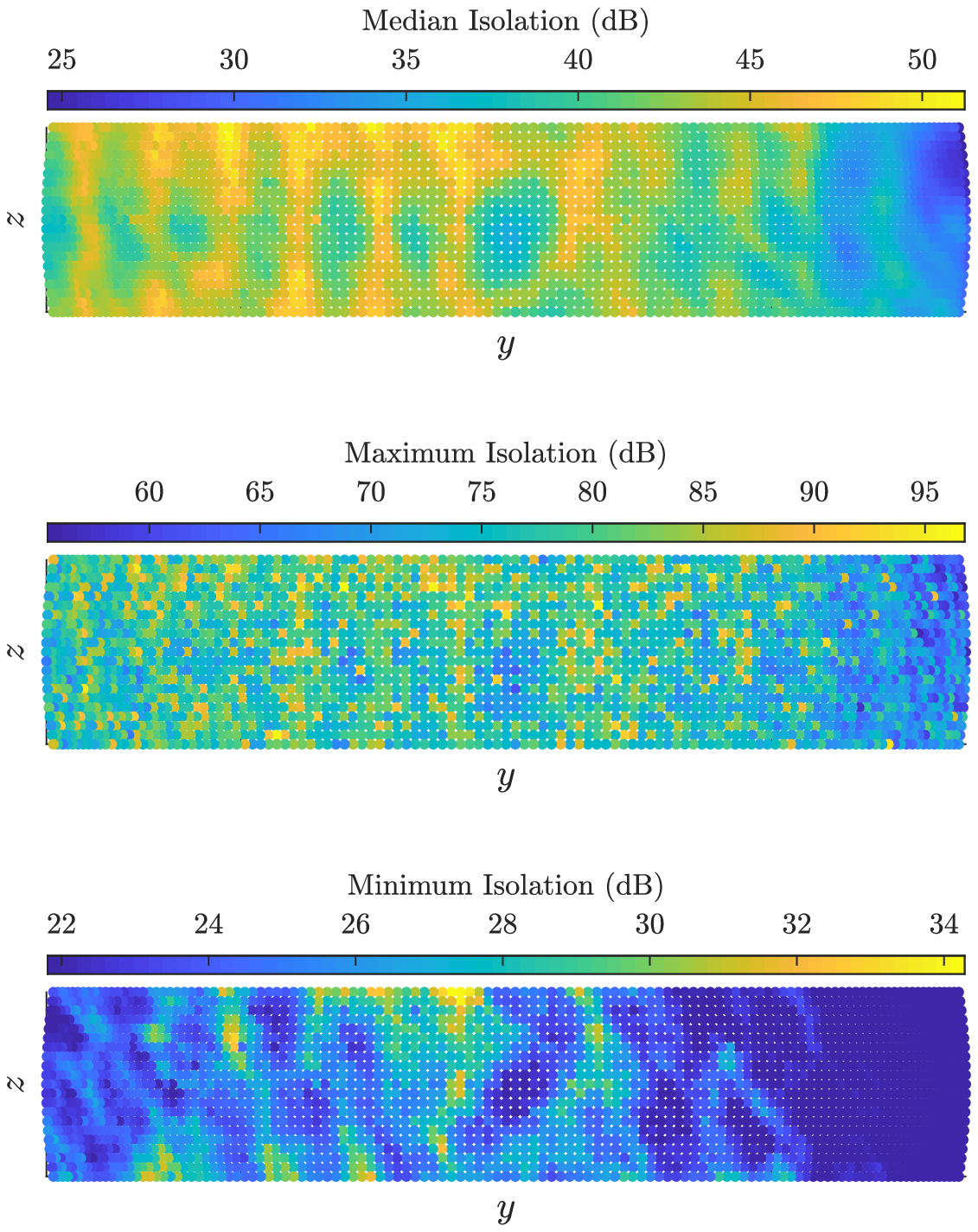}\label{fig:mean-median-max-min-per-tx-full}}
    \quad
    \subfloat[Observed by each receive beam.]{	\includegraphics[width=0.44\linewidth,height=\textheight,keepaspectratio]{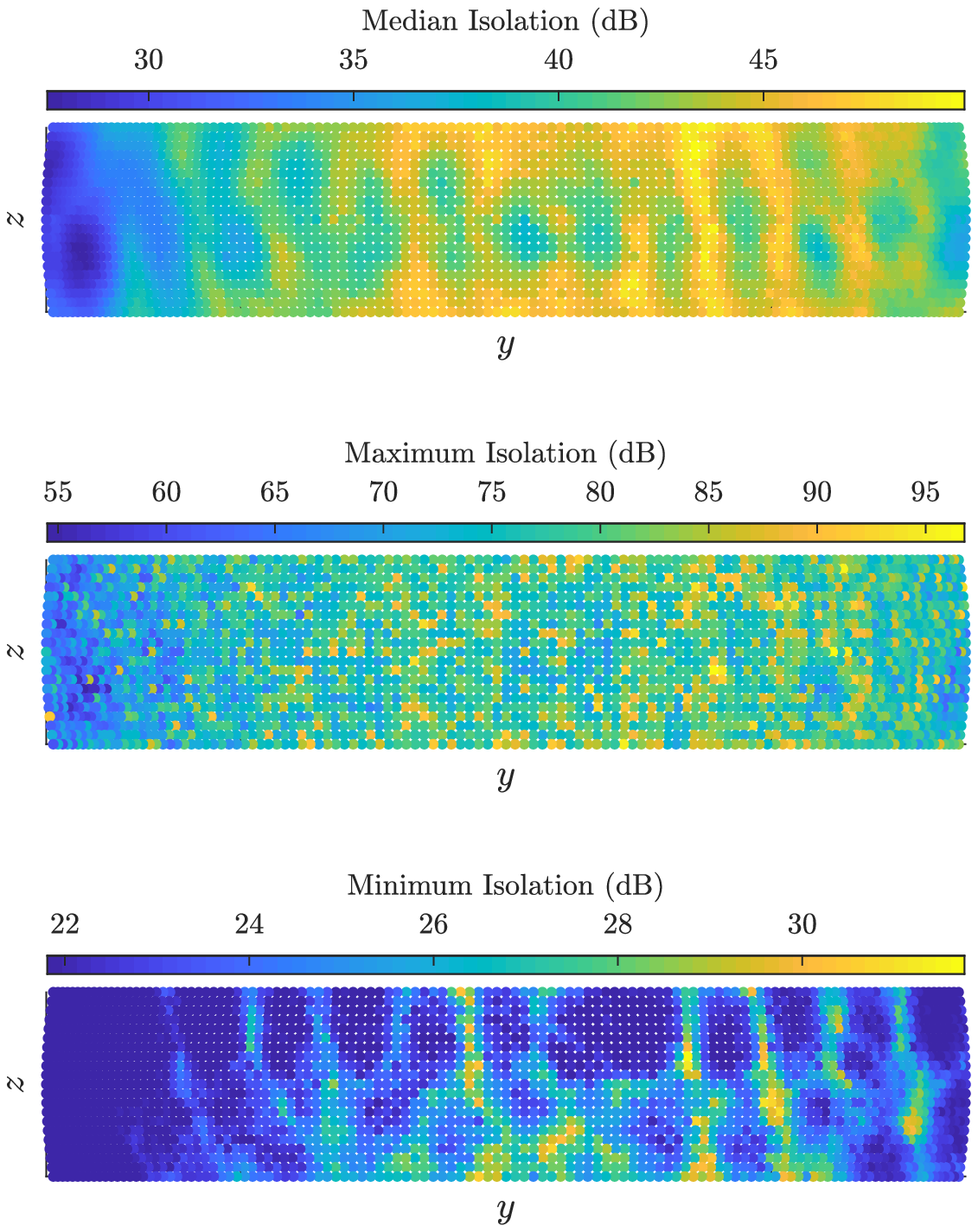}\label{fig:mean-median-max-min-per-rx-full}}
    \caption{For each transmit beam and receive beam, shown are the median, maximum, and minimum isolation across all receive and transmit beams, respectively.}
    \label{fig:mean-median-max-min-full}
\end{figure*}

\section{Spatial Characteristics} \label{sec:spatial}


We now hone in on narrower perspectives to better visualize and interpret our measurements.
First, let us begin by considering \figref{fig:mean-median-max-min-per-tx-full}, where each dot shows the median, maximum, and minimum isolation observed by each transmit beam across all receive beams; each dot is over a column in $\mL$.
\figref{fig:mean-median-max-min-per-rx-full} similarly shows these statistics observed when receiving a particular direction; each dot is over a row in $\mL$.
Referencing \figref{fig:mean-median-max-min-per-rx-full}, we see that the median isolation per receive beam ranges from around $25$ dB to around $50$ dB.
The maximum isolation observed per receive beam is at least around $55$ dB and at most over $95$ dB, while the minimum isolation is at least around $22$ dB and at most around $32$ dB.
We similarly observe these statistics for each transmit beam in \figref{fig:mean-median-max-min-per-tx-full}, which tell a similar story visually and numerically as the receive side.

\begin{remark}
As intuition may suggest based on \figref{fig:chamber}, the results of \figref{fig:mean-median-max-min-per-tx-full} and \figref{fig:mean-median-max-min-per-rx-full} indicate that (i) transmitting toward the receiver (to the right) tends to offer less isolation and (ii) receiving toward the transmitter (to the left) tends to offer less isolation.
\end{remark}

We notice some symmetry between \figref{fig:mean-median-max-min-per-tx-full} and \figref{fig:mean-median-max-min-per-rx-full}, especially in median isolation, which validates some degree of channel symmetry/reciprocity.
Also, we see much more variation across $y$ than $z$, suggesting that the azimuth of the steering direction plays a greater role than elevation, which one may expect since our transmitter and receiver are separated in azimuth but not in elevation.

We see that, even when steering our transmitter toward the receiver, there exists some receive beam(s) that offer $55$ dB or more of isolation.
Likewise, when steering our receiver toward the transmitter, there exists some transmit beam(s) that offer around $55$ dB or more of isolation.
This suggests that, while transmitting toward the receiver and receiving toward the transmitter \textit{generally} results in lower isolation, there exist receive beams and transmit beams that \textit{can} offer high isolation.
In a similar fashion, the minimum isolation exhibits that there also exist choices that can lead to low isolation.


While it may seem obvious that transmitting toward the receiver and receiving toward the transmitter would couple the most self-interference, it was not clear that this would be the case since the transmit and receive arrays exist in the near-field of one another.
The far-field distance of our arrays is approximately $2.4$ meters based on the rule-of-thumb $2D^2 / \lambda$; recall, our arrays are separated by approximately $30$ cm.
Operating in a near-field regime, the highly directional beams created by our $16 \times 16$ \upas are not necessarily ``highly directional'' from the perspective of one another \cite{roberts_wcm}.

\begin{remark}
    From this, we can conclude that there are not transmit beams nor receive beams that \textit{universally} offer high or low isolation---though there exist those that \textit{tend} to.
    Rather, the amount of isolation observed depends heavily on one's choice of transmit beam \textit{and} receive beam.
\end{remark}


%


\begin{figure*}
    \centering
    \subfloat[Observed by each transmit beam.]{\includegraphics[width=0.44\linewidth,height=\textheight,keepaspectratio]{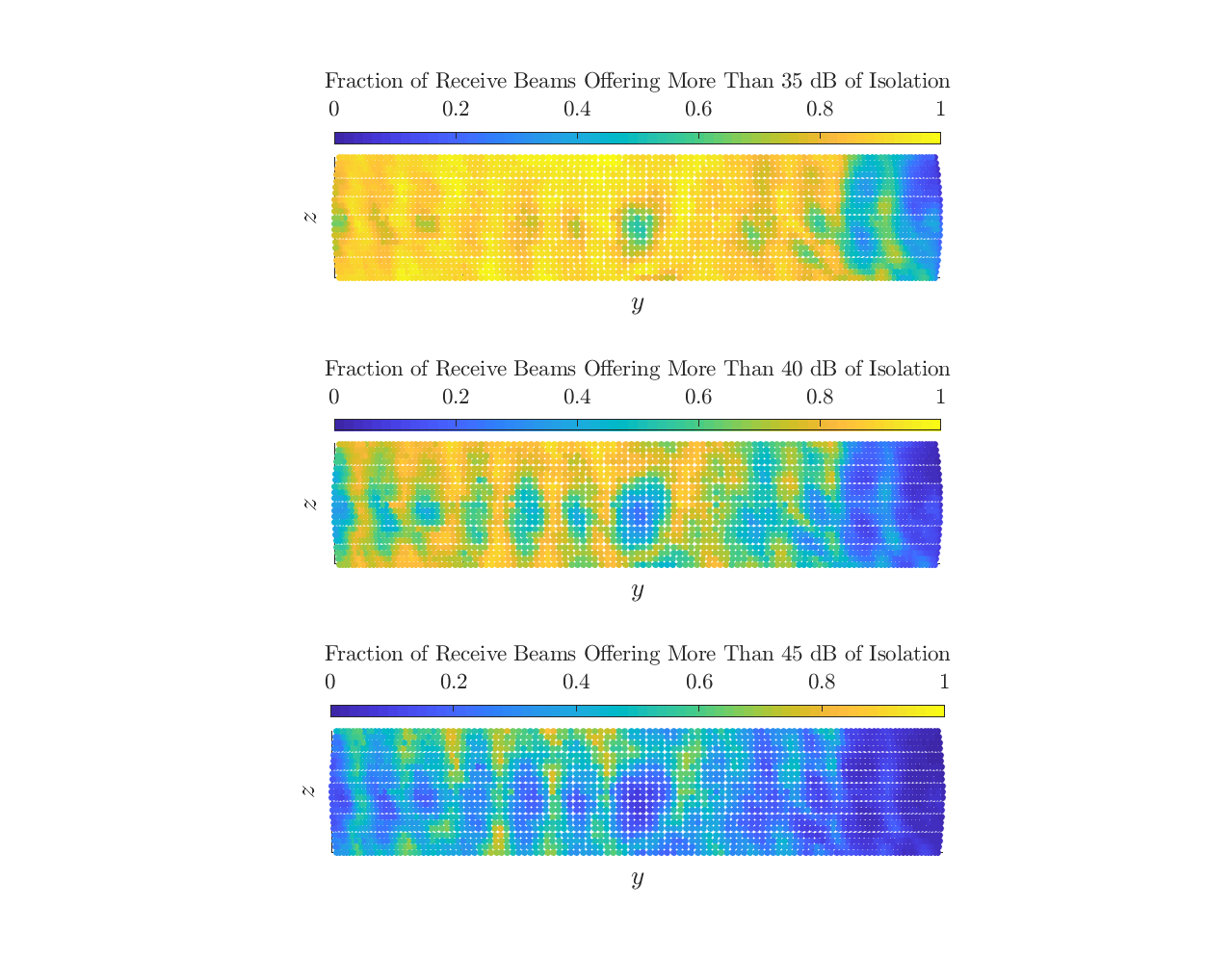}\label{fig:thresh-per-tx-full}}
    \quad
    \subfloat[Observed by each receive beam.]{	\includegraphics[width=0.44\linewidth,height=\textheight,keepaspectratio]{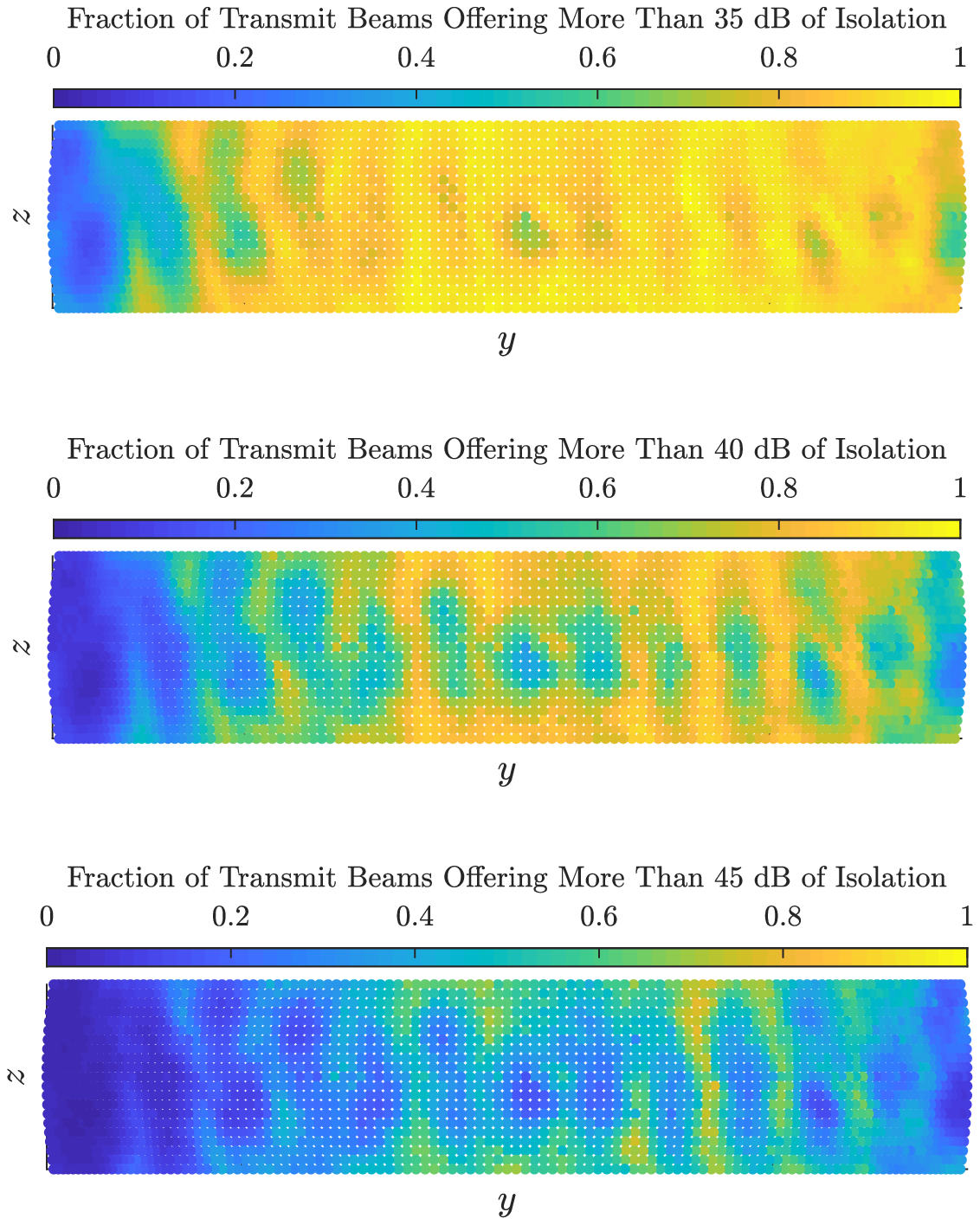}\label{fig:thresh-per-rx-full}}
    \caption{For each transmit beam and receive beam, the fraction of receive and transmit beams offering certain levels of isolation, respectively.}
    \label{fig:thresh-full}
\end{figure*}


In \figref{fig:thresh-per-tx-full}, for each transmit beam, we look at the fraction of receive beams that offer at least $35$, $40$, and $45$ dB of isolation.
Similarly, in \figref{fig:thresh-per-rx-full}, for each receive beam, we look at the fraction of transmit beams  that offer these same levels of isolation.
These two figures begin to further explain where the bulk of low isolation beam pairs come from: most of their density (visible in the \gls{cdf} in \figref{fig:dist-full-cdf}) lay in transmit beams steering toward the receive array and in receive beams steering toward the transmit array.
These directions correspond to the transmit and receive directions that couple high amounts of self-interference \textit{almost regardless} of the receive and transmit direction, respectively.
For example, when transmitting to the upper right, only around $10$--$20$\% of receive beams offer more than 35 dB of isolation, as evidenced by the top plot in \figref{fig:thresh-per-tx-full}.

Less dramatically, a large portion of the moderately high to high isolation beam pairs can be seen as coming from select transmit and receive directions.
In fact, these directions seem to be fairly agnostic to elevation, existing as vertical strips of bright dots. 
These vertical strips of high isolation separated by strips of low isolation are speculated to be caused by the interaction of side lobes (which are not well defined in the near-field).
{The transmit and receive beams offering high isolation across large fractions of receive beams and transmit beams, respectively, correspond to \textit{approximate} transmit and receive nulls at the channel input and output (i.e., approximate left and right null spaces of $\mH$), respectively, that offer high isolation.}
Recall, from \figref{fig:mean-median-max-min-full}, we did not see any transmit or receive beams that \textit{universally} provided high isolation.

%

From both figures, we see that transmitting and receiving in and around broadside (azimuth and elevation of zero) leads to low isolation across many receive beams and transmit beams, respectively.
These results are not necessarily expected nor easily explained; it can perhaps be attributed to the presence of side lobes and their enhancement in this near-field setting.

%
%



\begin{figure*}
    \centering
    \subfloat[Observed by each transmit beam for a given receive beam.]{\includegraphics[width=0.44\linewidth,height=\textheight,keepaspectratio]{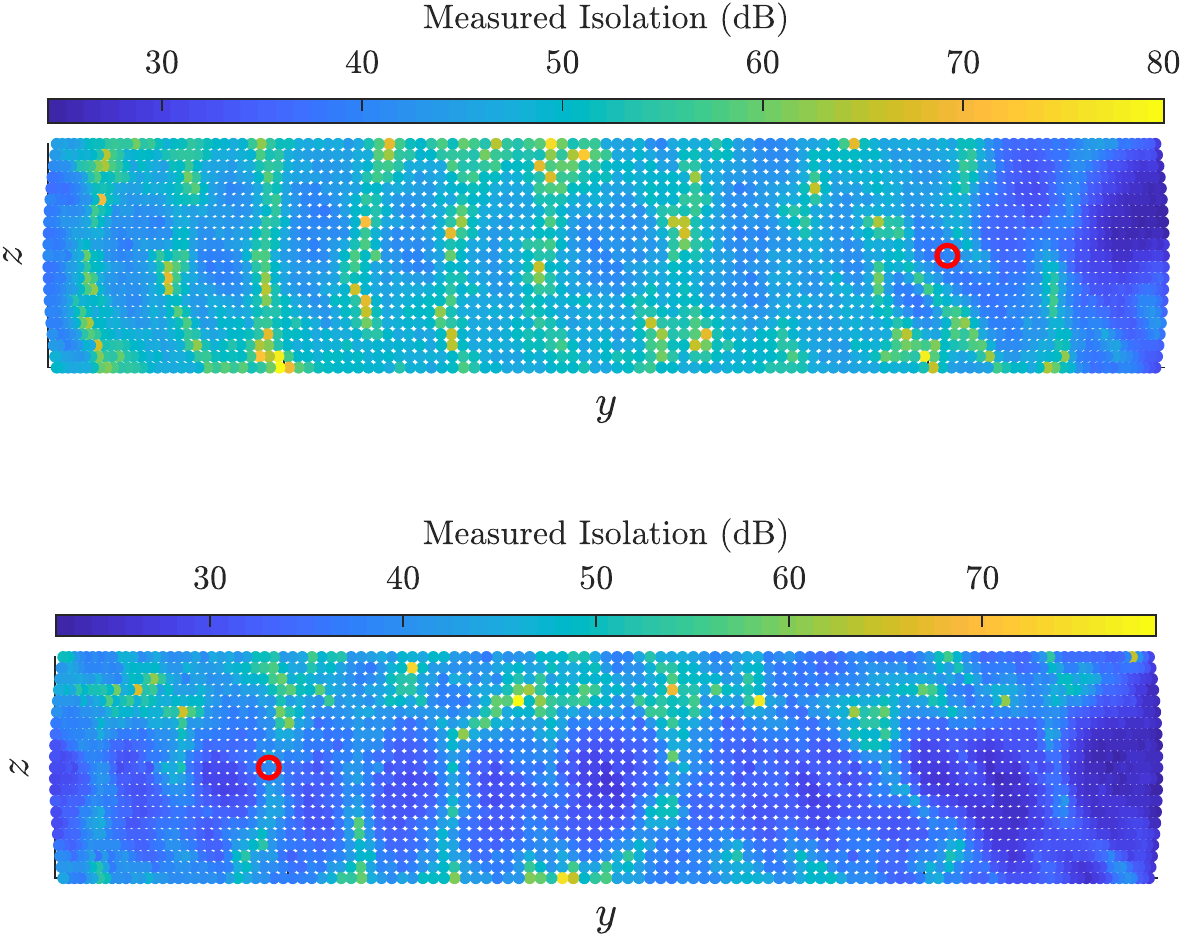}\label{fig:select-rx-per-tx-beam}}
    \quad
    \subfloat[Observed by each receive beam for a given transmit beam.]{	\includegraphics[width=0.44\linewidth,height=\textheight,keepaspectratio]{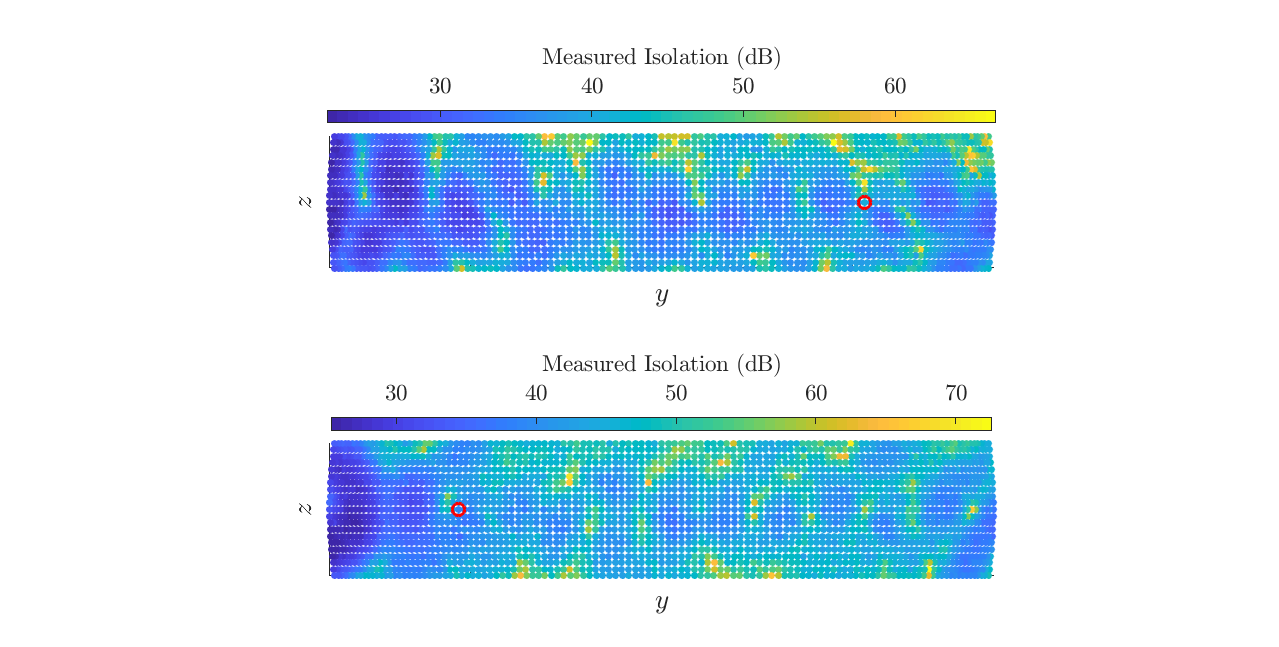}\label{fig:select-tx-per-rx-beam}}
    \caption{The isolation achieved across transmit and receive beams for particular receive and transmit beams (shown as red $\circ$), respectively.}
    \label{fig:select-full}
\end{figure*}

Honing in further, we now look at the isolation achieved at each transmit beam \textit{for a particular receive beam} and at each receive beam \textit{for a particular transmit beam}, as shown in \figref{fig:select-rx-per-tx-beam} and \figref{fig:select-tx-per-rx-beam}; this can be thought of as considering a single row or column of the matrix $\mL$, respectively.
Let us first consider the isolation observed across receive beams for particular transmit beams.
In \figref{fig:select-tx-per-rx-beam}, we have selected two transmit directions: toward the receive array and away from the receive array.
For each, have shown the isolation measured between the transmit beam and each receive beam.
As expected, when the transmit beam changes, the isolation profile across receive beams also changes.

When transmitting to the left (away from the receiver; bottom plot), we see that receive beams offer moderately high or high isolation more often.
Receiving toward the transmitter leads to low isolation, unavoidably.
When transmitting toward the receiver (top plot), receive beams offering moderately high to high isolation are less widespread. 
They exist, however, particularly in high and low elevation.
Also, we see a small vertical strip of higher isolation receive beams that steer leftward, toward the transmit array.
{This reinforces that isolation may \textit{tend} to be lower when transmitting toward the receiver and receiving toward the transmitter but is not \textit{universally} the case.}
Looking at both plots in \figref{fig:select-tx-per-rx-beam}, the receive beam that offers maximum isolation varies with transmit beam, which further backs our claim that there are not receive beams that universally offer high isolation.


Similarly, in \figref{fig:select-rx-per-tx-beam}, we have selected two receive directions and, for each, have shown the isolation measured between the receive beam and each transmit beam.
Analogous conclusions can be drawn as with \figref{fig:select-tx-per-rx-beam}, though there are useful comments to make.
Again, varying with each receive beam, there exists an isolation-maximizing transmit beam.
Notice that even when the receive beam is steered away from the transmit array (to the right; top plot), transmitting toward the receive array (to the right) still inflicts substantial self-interference.

We observe a certain degree of symmetry in \figref{fig:select-full}.
Transmitting toward the receiver (top \figref{fig:select-tx-per-rx-beam}) is similar to receiving toward the transmitter (bottom \figref{fig:select-rx-per-tx-beam}).
Transmitting away from the receiver (bottom \figref{fig:select-tx-per-rx-beam}) is similar to receiving away from the transmitter (top \figref{fig:select-rx-per-tx-beam}).
This further verifies a sense of symmetry of our self-interference channel $\mH$.

\begin{remark}
    We can clearly see that simply steering the transmitter away from the receiver \textit{or} steering the receiver away from the transmitter does not offer widespread isolation.
    {Moreover, the high-isolation receive directions are often quite narrow in the sense that small changes in receive direction can lead to significant changes in isolation.}
    For instance, when transmitting toward the right, the isolation across receive beams varies by about 45 dB, and we see that shifting a receive beam by only $1^\circ$--$2^\circ$ degrees in azimuth and/or elevation can lead to changes of $25$ dB in isolation or more.
    We note that this sensitivity to steering direction is much more apparent with high-isolation beams than low-isolation ones.
\end{remark}

%

\section{Conclusion} \label{sec:conclusion}
Inspection of the 28 GHz self-interference channel between transmit and receive arrays has shown that highly directional \mmwave beams \textit{can} offer high levels of isolation but are not necessarily likely to. 
From 6.5 million measurements, we observe that transmitting toward the receive array or receiving toward the transmit array {tends} to couple more self-interference.
Likewise, we see that transmitting away from the receive array or receiving away from the transmit array {tends} to couple less self-interference, but no transmit or receive beams are guaranteed to offer widespread high isolation.
We simultaneously observe that small changes in steering direction can lead to significant variability in the degree of self-interference coupled.
Valuable future work includes self-interference channel modeling, investigating the impact of environmental reflections, and the design of beamforming codebooks for \mmwave full-duplex.

\comment{
In this measurement campaign we have inspected the isolation levels afforded across a dense number of transmit and receive phased array beams coupled by a 28 GHz self-interference channel.
We observe that the isolation achieved with conventional, highly directional transmit and receive beams \textit{can} reach levels sufficient for full-duplex operation, though this is not guaranteed nor is it close to typical, as isolation across all $6.5$ million measured beam pairs ranges by more than $75$ dB.
This suggests that achieving the isolation levels sufficient for \mmwave full-duplex with beamforming alone will likely require strategically selecting transmit and receive beams.
Characterization of how isolation is spread around transmit-receive beam pairs indicates two important points: (i) steering directions that achieve high isolation are often quite narrow and (ii) low-isolation beam pairs often require above average deviation to observe increased isolation.
Both points emphasize that it requires finely measuring the self-interference channel and subsequently accurately steering beams to achieve isolation levels sufficient for high full-duplexing gain.
Valuable future work would be in self-interference channel estimation and modeling, investigating the impact of environmental reflections, and the design of beamforming codebooks.
}












\section*{Acknowledgments}

I.~P.~Roberts is supported by the National Science Foundation Graduate Research Fellowship Program under Grant No.~DGE-1610403. Any opinions, findings, and conclusions or recommendations expressed in this material are those of the author(s) and do not necessarily reflect the views of the National Science Foundation.


\bibliographystyle{bibtex/IEEEtran}
\bibliography{bibtex/IEEEabrv,refs}

\end{document}